\documentclass{aastex}

\shorttitle{LMC GLOBULAR CLUSTERS}
\shortauthors{van den Bergh}

\begin{document}

\title{WERE THE LMC GLOBULAR CLUSTERS FORMED IN A DISK?}

\author{Sidney van den Bergh}
\affil{Dominion~Astrophysical~Observatory, Herzberg~Institute~of~Astrophysics, National~Research~Council~of~Canada, 5071~W.~Saanich~Rd., Victoria,~British~Columbia, V9E~2E7, Canada}
\email{sidney.vandenbergh@nrc-cnrc.gc.ca}

\begin{abstract}
The radial velocities of the 13 globular clusters in the Large Magellanic Cloud have a dispersion of 28 km s$^{-1}$ relative to the HI rotation curve of the LMC, compared to a dispersion of 30 km s$^{-1}$ with regard to the mean globular cluster velocity. This shows that, contrary to a suggestion by Schommer et al. (1992), one cannot yet rule out the possibility that the LMC globular clusters formed in a pressure supported halo, rather than in rotating disk. The globular clusters in the LMC may therefore, after all, exhibit a relationship between age and kinematics that is similar to that of the clusters in M33. 
\end{abstract}


\section{INTRODUCTION}

     In a pioneering investigation of the clusters in the Large Magellanic Cloud, Freeman, Illingworfth \& Oemler (1983) surmised that ``[T]here is no old, kinematical halo population among the clusters of the LMC.'' At that time radial velocities were only available for nine old clusters in the Large Cloud. Presently 13 globular clusters are known to be associated with the Large Magellanic Cloud (Schommer 1991, Schommer et al. 1992, Suntzeff 1992, van den Bergh 2000, p. 104). From their discussion of the kinematics of old clusters in the LMC Schommer et al. (1992) concluded that: ``The oldest clusters still present an enigma; they do not have the kinematics of an isothermal, or slowly rotating, pressure-supported halo. These objects rotate with an amplitude comparable to that of the younger disk''. Taken at face value this conclusion would appear to indicate that the oldest clusters in the Large Cloud formed quite differently from those in M33 (Schommer et al. 1991, Chandar et al. 2002), in which the oldest (globular) clusters are observed to have halo kinematics, whereas younger (open) clusters exhibit disk-like motions. The view that late-type galaxies such as M33 (M V = -18.9) and the LMC (M V = -18.5), which have similar uminosities, had very different evolutionary histories would not fit comfortably with most current views of galaxy evolution. This prompts one to ask how secure is the conclusion that the oldest massive clusters in the Large Cloud formed in a disk, rather than in the LMC halo?

\section{KINEMATICS OF OLDEST LMC CLUSTERS}

     In their Table 2 Schommer et al. (1992) provide radial velocities (corrected for an Eastward transverse velocity of 150 km s$^{-1}$ for the 13 oldest LMC clusters which belong to SWB class VII (Searle, Wilkinson \& Bagnuolo 1980). These data are plotted in Figure 1, which shows that the rms velocity dispersion around the LMC HI rotation curve (thin line) is 28 km s $^{-1}$, compared to a dispersion of 30 km s$^{-1}$ around the mean velocity (dashed line) of the SWB VII clusters. This result shows that the H I rotation curve does not provide a significantly better fit to the the kinematics of the oldest LMC clusters than does a model in which these clusters are not assumed to be supported by rotation. In this respect the old clusters of class WWB VII differ significantly from the intermediate-age clusters of SWB classes IV, V and VI which (see Figure 5 of Schommer et al. 1992) clearly exhibit a significant rotational velocity component. It is interesting to note that the 30
$\pm$6 km s$^{-1}$ velocity dispersion of the 13 SWB VII clusters appears to be smaller, albeit only at the  2 $\sigma$ level, than the 53 $\pm$10 km s$^{-1}$ velocity dispersion that Minniti et al. (2003) have recently found for 42 RR Lyrae variables in the Large Cloud. However, it is noted that the velocity dispersion of the LMC RR Lyrae variables is comparable to the 54 $\pm$8 km s$^{-1}$ velocity dispersion that Chandar et al. obtain for the old M33 globular clusters with $R < 2.25$ kpc, but smaller than the 82 $\pm$13 km s$^{-1}$ dispersion that these authors find for the old M33 halo clusters with 
$R > 2.25$ kpc.

\section{CONCLUSIONS}

     By lumping old and intermediate-age LMC clusters together Schommer et al. (1992) found that these clusters exhibited significant rotation. However, the data plotted in Figure 1 show that the oldest (SWB class
VII) clusters do not exhibit a significant rotational component. In this respect the Large Cloud cluster system resembles M33 in which Chandar et al. find that 85 $\pm$ 5 percent of the oldest clusters have halo kinematics. It is therefore concluded that worries about the apparent difference between the kinematics of the oldest population components in M33 and in the LMC (van den Bergh 2000, p. 83) were probably premature.

     I thank Ken Freeman for pointing out to me that the number of old LMC globular clusters might not be large enough to prove the conclusion that these clusters have disk kinematics. Thanks are also due to Peter Stetson for statistical advice.

\clearpage

\figcaption{Plot of radial velocity (corrected for transverse motion) versus position angle for the 13 oldest LMC globular clusters, which belong to class SWB VII. The continuous curve shows the H I rotation curve for the inner part of the Large Cloud, and the dashed line plots the mean of the cluster radial velocities. The Figure shows that the rotation versus position angle curve and the mean cluster velocity provide equally good fits to the velocity data for the oldest LMC clusters.}

\end{document}